\documentclass[12pt]{article}
\usepackage{amsmath, amsfonts, amssymb, amsxtra, amsfonts, amsthm, bbold, natbib, subcaption, xcolor}
\usepackage{graphicx}
\usepackage[margin = 1in]{geometry}
\usepackage{multirow}
\usepackage{threeparttable}
\usepackage{booktabs}
\usepackage{amsmath}

\definecolor{dacolor}{RGB}{0,0,0} 
\newcommand{\bmath}[1]{\protect{\boldsymbol{#1}}}
\newcommand{\bbm}[1]{\protect{\mathbb{#1}}}
\newtheorem{assumption}{Assumption}

\newcommand{\dr}{\protect{d_r}}
\newcommand{\dt}{\protect{d_t}}
\newcommand{\calL}{\mathcal{L}}

\newcommand{\Rbar}{\bar{\mathbb{R}}}

\newcommand{\btheta}{\bmath{\theta}}
\newcommand{\bgamma}{\bmath{\gamma}}

\newcommand{\balpha}{\bmath{\alpha}}
\newcommand{\bkappa}{\bmath{\kappa}}
\newcommand{\bbeta}{\bmath{\beta}}
\newcommand{\bx}{\bmath{x}}
\newcommand{\bX}{\bmath{X}}
\newcommand{\gi}{\protect{\gamma_{i}}}
\newcommand{\dri}{\protect{\delta^{R}_{i}}}
\newcommand{\dti}{\protect{\delta^{T}_{i}}}
\newcommand{\yri}{\protect{y^{R}_{i}}}
\newcommand{\yti}{\protect{y^{T}_{i}}}
\newcommand{\Yri}{\protect{Y^{R}_{i}}}
\newcommand{\Yti}{\protect{Y^{T}_{i}}}
\newcommand{\expb}[1]{\exp\left\{ #1 \right\}}
\newcommand{\sigmainv}{{\sigma^{-1}}}
\newcommand{\bthetab}{\btheta^{(b)}}
\newcommand{\gib}{\gi^{(b)}}
\newcommand{\cindep}{\mathrel{\text{\scalebox{1.07}{$\perp\mkern-10mu\perp$}}}}

\title{Survivor average causal effects for continuous time: a principal stratification approach to causal inference with semicompeting risks}

\author{Leah Comment, Fabrizia Mealli, Sebastien Haneuse, and Corwin Zigler}

\begin{document}


\maketitle







\label{firstpage}


\begin{abstract}
In semicompeting risks problems, nonterminal time-to-event outcomes such as time to hospital readmission are subject to truncation by death. These settings are often modeled with illness-death models for the hazards of the terminal and nonterminal events, but evaluating causal treatment effects with hazard models is problematic due to conditioning on survival-- a post-treatment outcome-- that is embedded in the definition of a hazard. Extending an existing survivor average causal effect (SACE) estimand, we frame the evaluation of treatment effects in the context of semicompeting risks with principal stratification and introduce two new causal estimands: the time-varying survivor average causal effect (TV-SACE) and the restricted mean survivor average causal effect (RM-SACE). These principal causal effects are defined among units that would survive regardless of assigned treatment. We adopt a Bayesian estimation procedure that parameterizes illness-death models for both treatment arms. We outline a frailty specification that can accommodate within-person correlation between nonterminal and terminal event times, and we discuss potential avenues for adding model flexibility. The method is demonstrated in the context of hospital readmission among late-stage pancreatic cancer patients.
\end{abstract}

%
\newbox\keywdbox
\def\keywords{\global\setbox\keywdbox\vbox\bgroup\hsize\textwidth\small\leftskip0pc\rightskip\leftskip\noindent{\sc
Key words:\hskip1em}\ignorespaces}
\def\endkeywords{\egroup}




%

\section{Introduction}
\label{sec:introduction}

Survival remains the gold standard for evaluating treatments in high-mortality settings, but doctors, patients, and policymakers also make decisions based on outcomes related to quality of the remaining lifespan, e.g., hospitalization, onset of dementia, or loss of independence. Comparing treatments on the basis of non-mortality time-to-event outcomes is complicated by the fact that interventions on quality of life can also affect mortality. Death is a ``truncating'' or ``terminal'' event in that it precludes the occurrence of the ``nonterminal'' event. As a motivating example, consider a hospital tasked with reducing readmissions among late-stage cancer patients. Death is a terminal truncating event because its occurrence precludes future hospitalizations, and a hospital readmission is a nonterminal event which does not truncate death. Imagine an intervention which increases every individual's risk of readmission while simultaneously harming survival among the patients most at risk for readmission; a naive analysis may (unfairly) conclude the intervention \emph{reduces} readmissions on the basis that fewer patients in that group survive long enough to experience readmission. This danger stems from a problem known as ``truncation by death'' and has been addressed in the causal inference literature using principal stratification \citep{zhang_estimation_2003}.

Principal stratification handles truncation by death by defining causal contrasts to be restricted among the group that would not experience the truncating event under either treatment at a fixed time point. Initially introduced by \citet{robins_new_1986} and formalized in \citet{rubin_causal_2000} and \citet{zhang_estimation_2003}, the traditional survivor average causal effect (SACE) is the causal effect of the treatment on the truncated (i.e., non-mortality) outcome among the subpopulation that would survive regardless of treatment assignment. A number of papers have discussed the nonparametric identifiability conditions and assumptions for the SACE \citep{long_sharpening_2013,zhang_estimation_2003,robins_new_1986,tchetgen_identification_2014}. 
The time point for defining always-survivorship is often implicit, such as ``by the end of the study,'' and typically only one such time $t$ is considered. With a time-to-event structure of both the terminal (truncating) event and the nonterminal event, explicit definition of causal effects is indexed by both: (1) the time defining the ``always survivors,'' denoted with $t$, and (2) the time interval over which treatment contrasts on the nonterminal event are evaluated, $r$. For example, interest may lie in the causal effect on cumulative incidence of  readmission at 30 days post-discharge ($r=30$) among patients who would survive under either treatment at 60 days ($t=60$). Examining such quantities across different values of $(r,t)$ can serve different inferential purposes. We use the term ``snapshot causal effect'' to describe survivor causal effects with $r=t$, where the same point in time is used in the definition of an ``always''-survivor (i.e., $t$) and the end point of the interval $(0,r]$ over which treatment effects on the non-mortality outcome are evaluated.

When the truncated outcome is time-to-event, estimating the SACE at a single $t$ can be problematic. First, providing one snapshot effect does not give decisionmakers information about the sensitivity of conclusions to the (possibly arbitrary) choice of $t$. If an intervention has different short- and long-term impacts, snapshot effects will provide a mixed or incomplete picture. More importantly, they do not account for the fact that \emph{timing} matters for the nonterminal event. With hospital readmission, being hospitalized earlier may lead to more total hospitalizations. Hospitalization could also accelerate death, if the post-readmission risk of death is higher. In other contexts -- such as the onset of dementia -- an earlier occurrence of the nonterminal event means more time spent in an unfavorable state, even if total lifespan remains unaffected. These concerns motivate the development of principal stratification methods that explicitly account for the time-to-event nature of the nonterminal outcome. 

Others have partially grappled with principal stratifications defined over time. For example, methods exist for treatment noncompliance in longitudinal contexts \citep{lin_longitudinal_2008,dai_partially_2012}. But unlike treatment compliance status, which can vary over time arbitrarily, death at $t$ under either treatment condition necessarily precludes membership in an always-alive state at $t' > t$. Defining strata on the basis of survival is also closely linked to principal strata generated by other continuously-scaled quantities \citep{schwartz_bayesian_2011}. Continuous variables can, in principle, create an infinite number of groups. Collapsing these into meaningful subpopulations for principal stratum causal effects is difficult and further complicated by problems of partial identifiability. Examples of continuous variables used to define principal strata include continuous measures of compliance \citep{jin_principal_2008, bartolucci_modeling_2011} and distance from a treatment location \citep{frangakis_principal_2007}.

In the survival analysis literature, the problem of nonterminal time-to-event outcomes which may be truncated by terminal events is referred to as semicompeting risks because the terminal event acts as a competing risk for the nonterminal event, but the reverse is not true\citep{fine_semi-competing_2001}. Models which accommodate this semicompeting risks structure have been applied to a wide range of settings, including hospital readmission \citep{lee_bayesian_2015}, cancer recurrence \citep{xu_statistical_2010}
, career advancement \citep{pan_estimating_2013}, and subscription product upgrades \citep{chen_hazards_2017}. Shared subject- and/or cluster-specific random effects, termed ``frailties,'' allow for correlation between event times that is induced by unmeasured factors \citep{xu_statistical_2010}. Such models are typically constructed on the hazard scale and account for truncation by removing individuals from nonterminal risk sets after the time of their observed terminal event; this removal is akin to what occurs with cause-specific hazards in competing risks problems. Joint modelling of the time-to-event outcomes is used to describe cumulative incidences or hazard-based predictive models \citep{lee_bayesian_2015}. Analyses of semicompeting risks emanating from the survival analysis literature typically do not focus explicitly on causal inference for treatment effects. Instead, these analyses estimate hazard ratios, which suffer from known limitations when the goal is causal inference \citep{hernan_hazards_2010}.

This paper addresses a gap in the literature by adapting existing semicompeting risks models and anchoring them to a principal stratification framework for the purpose of drawing causal inferences. We make four main contributions to the existing literature. First, we propose a framework for principal strata defined by a continuous time-to-event truncating variable, such as death time. Second, we motivate and define two new causal estimands for truncated time-to-event outcomes. Third, we describe a density factorization which is innovative for principal stratification problems and that allows for explicit links to (non-causal) semicompeting risks models. Lastly, a Bayesian estimation procedure is provided with accompanying software.


\section{A potential outcomes approach for semicompeting risks data}
\label{sec:poapproach}
\subsection{Notation}
Consider the evaluation of a binary intervention $Z$ (0=control, 1=treated), where interest lies in its effect on the times to a nonterminal event, $R$, and a terminal event, $T$, the occurrence of which may leave $R$ ill defined. We continue with the motivating setting of late-stage cancer care, where $Z$ is an intervention intended to reduce hospital readmission among recently discharged patients, $R$ is the time to hospital readmission, and $T$ is the time to death. The occurrence of death leaves future readmission undefined. Using the potential outcomes framework, let $R_i(z)$ and $T_i(z)$ denote the potential event times for readmission and death for person $i$, respectively, that would occur if the person were treated with $Z=z$. One or both of these events may be right censored by the potential censoring time $C_i(z)$. If death occurs without readmission, we set $R_i(z)$ to be $\bar{\mathbb{R}}$, a non-real value. The observed times are $Y_i^R = \min(R_i(Z_i), T_i(Z_i), C_i(z))$ and $Y_i^T = \min(T_i(Z_i), C_i(z))$, where $\min \left(\bar{\mathbb{R}}, x\right)$ is defined to be $x$ for any real $x$. The nonterminal event indicator $\delta^R_{i} = \bbm{1}\left(Y_i^R = R_i(Z_i)\right)$ is one if the nonterminal event is observed to occur and zero otherwise. The analogous death event indicator is $\delta^T_{i} = \bbm{1}(Y_i^T = T_i(Z_i))$. The set of covariates available at baseline, denoted by $\bX$, may consist of confounders, predictors of censoring, and measured baseline predictors of either event type. Together, the observed data for individual $i$ is $O_i = (\Yri, \dri, \Yti, \dti, \bX_i, Z_i)$.

\subsubsection{Principal stratification for continuous time}
A principal stratification is a partition of the population into subpopulations defined by joint values of the potential outcomes under all treatment conditions. Our basic principal strata are defined by the pair of potential death times $\left( T_i(0), T_i(1)\right)$. Since potential outcomes are not affected by treatment, stratifications based on the basic principal strata -- and unions of these strata -- exist prior to treatment assignment and can play a role similar to covariates. While the basic principal strata describe a unit's survival experience under both treatments across the entire time scale, it is useful to derive related quantities. For any $t$, let $V_i(t)$ denote the time-varying principal state \citep{lin_longitudinal_2008,dai_partially_2012} implied by the basic principal strata:
\begin{equation*}
V_i(t) = 
\begin{cases} 
AA & \text{if }T_i(0) > t, T_i(1) > t \\
TK & \text{if }T_i(0) > t, T_i(1) \le t \\
CK & \text{if }T_i(0)\le t, T_i(1) > t \\
DD & \text{if }T_i(0) \le t, T_i(1) \le t.
\end{cases}
\label{eqn:videf}    
\end{equation*}
The value of $V_i(t)$ represents a union of basic principal strata, depending on whether the individual is alive at $t$ in both arms ($AA$), alive only under treatment ($CK$) or control ($TK$), or dead under both ($DD$). In the context of hospital readmission for cancer patients, we may be interested in readmission differences among the ``always-alive'' at 30 days, i.e., $\{i : V_i(30) = AA \}$, as well as the net difference in 30-day survival probabilities $P\left(V(30) = TK\right) - P\left(V(30) = CK\right)$. 

The set of individuals with $V_i(t) = AA$ can also be viewed as a cohort with a well-defined and time-varying nonterminal event causal contrast function on interval $(0,t)$. For various $t$, we can define \emph{survivorship cohorts}, denoted by $\mathcal{A}_t$:
\[ \mathcal{A}_t = \{i: \min\left(T_i(0), T_i(1) \right) > t \}  = \{i: V_i(t) = AA \} \]
We note that $\mathcal{A}_{t'} \subseteq \mathcal{A}_t$ for $t' > t$. In the context of hospital readmission, $\mathcal{A}_{90}$ refers to the cohort of patients who would survive at least 90 days regardless of treatment assignment. Like the principal states, these principal strata are defined solely in terms of potential \emph{terminal} event times. Within a cohort $\mathcal{A}_t$, there can be no treatment effect on survival during the interval $(0, t)$; this fact ensures the time at risk for the nonterminal event is the same under both treatment and control conditions.

\subsection{Causal estimands for semicompeting risks} \label{sec:defineestimands}

\subsubsection{The time-varying survivor average causal effect (TV-SACE)}
On the cumulative incidence scale, the existing ``snapshot'' survivor average causal effect is
\begin{align}
SACE(t) = 
&
P\left(R(1) < t | V(t) = AA \right) -
P\left(R(0) < t | V(t) = AA \right).
\label{eqn:pointsace}
\end{align} 
As previously discussed, snapshot estimands do not describe time-varying effects for any well-defined population. When they are estimated at a single time point, as is typically done, it is also unclear how sensitive conclusions are to the choice of $t$. To address these limitations, we define a new quantity, the time-varying survivor average causal effect (TV-SACE). This estimand is a function taking two arguments $r$ and $t$, and it conveys the difference in the cumulative incidence of nonterminal events by time $r$ among the group that survives past $t > r$ regardless of assigned treatment:
\begin{align}
TV\text{-}SACE(r, t) = 
P\left(R(1) < r | V(t) = AA \right) -
P\left(R(0) < r | V(t) = AA \right).
\label{eqn:tvsace}
\end{align}
The TV-SACE captures the causal effect of $Z$ on $R$ that has manifested by time $r$, among the always-survivors at $t$. For example, hospitals may be interested in comparing 30-day ($r=30$) and 90-day ($r=90$) readmission rates among the cancer patients always-surviving at least 90 days post-discharge $(t=90)$. When $r = t$, the $TV\text{-}SACE(r,t)$ of Equation~\ref{eqn:tvsace} coincides with the $SACE(t)$ as defined in Equation~\ref{eqn:pointsace}.

The joint indexing of $TV\text{-}SACE(r, t)$ by both $r$ and $t$ is essential for characterizing causal effects. For a fixed $t$, the function $TV\text{-}SACE(r,t)$ (as a function of $r$) is a time-varying causal effect within the $\mathcal{A}_t$ cohort. It describes the accumulation of benefit causally attributable to treatment among the well-defined -- if latent -- cohort. The shapes of these curves for different $t$ reveal whether treatment effects steadily accrue or decay with time within the cohorts, and describe how the effect on the nonterminal outcome varies across subpopulations with different underlying risks of death.

Importantly, viewing $TV\text{-}SACE(r, t)$ as a function of $t$ does not characterize a time-varying causal effect, but a function of snapshot effects, each defined within a different $\mathcal{A}_t$ cohort. It does not represent a causal contrast varying over time in any static population, but the shape of the function $Q(t) = TV\text{-}SACE(t,t)$ has implications for study design as well as for the interpretation of results from any single study. If the population's underlying $Q(t)$ is believed to take on substantively different values over a range of relevant $t$, any snapshot $Q(t)$ gives an incomplete picture, and researchers should plan to estimate time-varying effects. Estimates of $Q(t)$ also give important information about the expected consistency of conclusions from studies estimating snapshot effects at different times. In essence, $Q(t)$ captures the sensitivity of causal effect estimation -- in both sign and magnitude -- to the moment in time used to define the always-survivorship group.

\subsubsection{The restricted mean survivor average causal effect (RM-SACE)}
Another causal estimand is a variation of the the restricted mean survival time (RMST) and captures the length of the delay in the nonterminal event among always-survivors. This effect may be particularly relevant if the nonterminal event represents a permanent state change, such as the onset of irreversible dementia.
\begin{align*}
RM\text{-}SACE(r, t) = 
\mathbb{E}\left[ \min(R(1), r) | V(t) = AA \right] -
\mathbb{E}\left[ \min(R(0), r) |V(t) = AA \right] 
\end{align*}

In the context of preventing hospital readmission, the $RM\text{-}SACE(r, t)$ captures how much expected hospitalization-free time the treatment causes one to accumulate by time $r$, defined among the always-survivors $\mathcal{A}_t$. Within the cohort $\mathcal{A}_t$, $RM\text{-}SACE(r,t)$ describes the timing of benefit accrual. If, within $\mathcal{A}_t$, the effect of treatment on the nonterminal events arises solely by delaying early events, the benefit accrues quickly and $RM\text{-}SACE(r,t)$ eventually levels off as $r$ increases. If different survivorship cohorts have dramatically different curves, then the effect on the nonterminal event is heterogeneous with respect to the underlying risk of death.

Just as with $Q(t)$, the function $M(t) = RM\text{-}SACE(t,t)$ conveys the sensitivity of the snapshot version to the choice of $t$. If $M(t)$ increases steadily, the choice of $t$ matters greatly, and reporting $RM\text{-}SACE(t,t)$ for only a single time point understates the total impact of the treatment on delaying the nonterminal event. On the other hand, if $M(t)$ levels off at some $t^*$, then any benefits attributable to treatment can be fully captured by an estimate of $M(t^*)$. 

\subsection{Structural assumptions}\label{sec:structural}
We now review a set of assumptions essential to our estimation strategy for $TV\text{-}SACE(r,t)$ and $RM\text{-}SACE(r,t)$. For clarity, our exposition focuses on non-recurrent nonterminal events, where any individual who experiences the nonterminal event is no longer at risk for that event. For nonterminal events which are non-permanent and that in principal could recur -- like a second hospital readmission -- the proposed framework may still be relevant with careful definition of the nonterminal event (e.g., time to first readmission). 

\begin{assumption}{Consistency of potential outcomes.}\label{ass:consistency}
\begin{align*}
R_i = & Z_i R_i(1) + (1 - Z_i) R_i(0) \\
T_i = & Z_i T_i(1) + (1 - Z_i) T_i(0)
\end{align*}
\end{assumption}
Consistency is a standard assumption throughout the causal inference literature which connects observables $R$ and $T$ to their corresponding potential outcomes. Briefly, the treatment is well-defined such that there are no hidden variations within treatment level \citep{rubin_comment:_1990}.

\begin{assumption}{Conditional exchangeability (no unmeasured confounding).}\label{ass:unconfoundedness} \\
The observed treatment assignment does not depend on the potential outcomes after accounting for the set of measured covariates $\bX$.
\begin{equation*}
\left( R(z), T(z) \right) \cindep Z \ \big\vert \  \bX \text{ for }z \in \{ 0,1 \}
\end{equation*}
\end{assumption}
In a randomized trial, this assumption holds by design since treatment assignment is independent of all measured and unmeasured variables. For observational settings, interpreting effect estimates as causal effects requires a sufficiently comprehensive $\bX$.

\begin{assumption}{Shared, non-informative censoring of event times.}\label{ass:noninformative}\\
The potential censoring times are shared (i.e., $C_i \equiv C_i(0) = C_i(1)$). Furthermore, the vector of potential censoring times $C$ is conditionally independent of all potential event times.
\begin{equation*}
\left( R(0), T(0), R(1), T(1)\right) \cindep C \big\vert \bX
\end{equation*}
\end{assumption}
Non-informative censoring is required for the consistent estimation of cumulative distribution functions. With administrative censoring, this assumption is satisfied by design.

\subsection{Connection to traditional semicompeting risks models}
We state a key simplifying assumption that builds a bridge to the semicompeting risks literature. With closely related nonterminal and terminal event processes, it is unrealistic to assume that any measured baseline set $\bX$ will contain all sources of dependence between potential event times in $\left(R(0), R(1), T(0), T(1) \right)$. However, if the cause of the dependence is baseline heterogeneity in the patient population, it may be reasonable to assume that baseline factors can be summarized by a one-dimensional subject-specific latent trait $\gamma_i$. As with any random effect, $\gamma_i$ cannot adjust for unmeasured confounding. However, $\gamma_i$ can be used to model sources of dependence in event times across treatment arms which are independent of the treatment assignment mechanism (i.e., unmeasured predictors).

\begin{assumption}{Independence of potential outcomes conditional on covariates and latent frailty.} \label{ass:cindepgivenfrailty} \\
Potential nonterminal and terminal event times under each treatment are conditionally independent conditional on $\bX$ and an individual-level latent trait $\gamma$.
\begin{equation*}
\left( R(0), T(0) \right) \cindep \left( R(1), T(1) \right) | \gamma, \bX
\end{equation*}
\end{assumption}

Assumption \ref{ass:cindepgivenfrailty} suggests a factorization of the joint density of the four potential outcomes $R(0)$, $R(1)$, $T(0)$, and $T(1)$ that is unusual within the principal stratification literature. Traditional model-based principal stratification approaches build a model for stratum membership given covariates (the ``S-model''), and a model for the joint distribution of the potential outcomes conditional on the principal strata and covariates (the ``Y-model'') \citep{schwartz_bayesian_2011}. Instead, we choose an alternative factorization, shown in Equation~\ref{eqn:rtfactor}, which further simplifies to Equation~\ref{eqn:rtfactorsimplified} under Assumption \ref{ass:cindepgivenfrailty}.
\begin{align}
f\left(R(0), R(1), T(0), T(1) | \bX, \gamma \right)
= &
f\left(R(0), T(0) | \bX, \gamma \right) f\left(R(1), T(1) | R(0), T(0), \bX, \gamma \right) \label{eqn:rtfactor} \\
= & f\left(R(0), T(0) | \bX, \gamma \right) f\left(R(1), T(1) | \bX, \gamma \right) \label{eqn:rtfactorsimplified}
\end{align}
This arrangement makes it easy to enforce that $T_i(z)$ must exceed $R_i(z)$ whenever the nonterminal event occurs (i.e., $R_i(z) \in \mathbb{R}^+$). We can also leverage existing illness-death transition models from the semicompeting risks literature to obtain a general form of the likelihood. 

\subsection{Likelihood}\label{sec:likelihood}
Within a single treatment condition, the semicompeting risks structure of the potential outcomes $R(z)$ and $T(z)$ can be seen as an illness-death transition model characterizing transitions among the event-free (``healthy''), nonterminal only (``ill''), and post-terminal (``dead'') states. Hazards can be defined for the three types of event transitions: (1) healthy-ill, (2), healthy-dead, and (3) ill-dead.
\begin{align*}
\lambda_1^z(r) 
& = 
\lim_{\Delta \to 0} \frac{P\left(Y^R(z) \in [r, r + \Delta) | Y^R(z) \ge r, Y^T(z) \ge r \right)}{\Delta}\\
\lambda_2^z(t) 
& = 
\lim_{\Delta \to 0} \frac{P\left(Y^T(z) \in [t, t + \Delta) | Y^R(z) \ge t, Y^T(z) \ge t \right)}{\Delta} \\
\lambda_2^z(t | r) 
& = 
\lim_{\Delta \to 0} \frac{P\left(Y^T(z) \in [t, t + \Delta) | Y^R(z) = r, Y^T(z) \ge r \right)}{\Delta}
\end{align*}

The treatment arm-specific hazards conditional on covariates are denoted $\lambda_1^z(t|\bx_i, \gamma_i, \btheta)$, $\lambda_2^z(t|\bx_i, \gamma_i, \btheta)$, and $\lambda_3^z(t|r, \bx_i, \gamma_i, \btheta)$, where $\btheta$ is a vector of unknown parameters. With cumulative hazard $\Lambda_j^z(t|\cdot)=\int_0^t \lambda_j^z(u|\cdot) du$, the observed data likelihood conditional on $O_i$ is given by
\begin{align}
\mathcal{L}_{c}
= \prod_{i=1}^n \bigg( 
&
[\lambda^{Z_i}_{1}(\yri | \bx_i, \gamma_i, \btheta)]^{\dri}
[\lambda^{Z_i}_{2}(\yri | \bx_i, \gamma_i, \btheta)]^{\dti(1-\dri)}
[\lambda^{Z_i}_{3}(\yti | \yri, \bx_i, \gamma_i, \btheta)]^{\dti\dri}
\nonumber \\
& \times 
\expb{ -
\Lambda^{Z_i}_{1}(\yri | \bx_i, \gamma_i, \btheta) -
\Lambda^{Z_i}_{2}(\yri | \bx_i, \gamma_i, \btheta) - 
\Lambda^{Z_i}_{3}(\yti | \yri, \bx_i, \gamma_i, \btheta)} \bigg)
\end{align}
If the frailties are included as unknown parameters in an expanded parameter set $\btheta^* = (\btheta, \bgamma)$ for $\bgamma = (\gamma_1,\dots,\gamma_n)'$, the dimension of the parameter space is large and grows linearly with $n$, rendering estimation impracticable for large data sets. For computational efficiency and scalability, we use the marginalized likelihood $\mathcal{L}_{m} = \int \mathcal{L}_{c} f(\bgamma) d\bgamma$ rather than the conditional likelihood in our estimation algorithm. For selected choices of $f(\bgamma)$, the form of $\mathcal{L}_m$ can be obtained analytically, but numerical integration within the MCMC can be used to accommodate arbitrary $f(\bgamma)$. Computationally feasible estimation strategies are the focus of the next section.

\section{Bayesian model-based estimation of causal effects}\label{sec:connectionscr}
\subsection{Identifiability in the Bayesian framework}

We propose a Bayesian approach anchored to illness-death models for state transitions. Note that the likelihood in Section~\ref{sec:likelihood} does not support point identifiability of the principal stratum causal effects, a problem which also arises with the more traditional (i.e., snapshot) SACE \citep{long_sharpening_2013}. This motivates our use of a Bayesian estimation procedure. In addition to the ability to handle large amount of missing data (including unobserved potential outcomes) in much the same way as unknown parameters, the Bayesian procedure with proper prior distributions will yield proper posterior inference, even in the face of flat portions of the likelihood. In these instances, some of the unknown parameters in $\btheta$ are only ``partially identified'': even with infinite amounts of data, the posterior distribution converges to a non-degenerate distribution over a range of possible values that is smaller than that specified in the prior, but not equal to a single point \citep{gustafson_bayesian_2010}. 

\subsection{Implementation with parametric illness-death models} \label{sec:modelspec}
In this paper we focus on hazards parameterized using Weibull regression models for each of the six possible transitions. Although alternative specifications are possible, we elect to use a semi-Markov model for the terminal event after the occurrence of the nonterminal event (i.e., for $t > R_i(z)$, the terminal event hazard at $t$ depends on $R_i(z)$ only through $(t - R_i(z))$ \citep{lee_bayesian_2015}. For $z \in \{ 0, 1\}$ and $j \in \{1,2,3\}$, the Weibull shape for transition $j$ under $Z=z$ is denoted $\alpha^z_j$, and the baseline hazard rate is $\kappa^z_j$, giving hazard equations:
\begin{align*}
\lambda^z_{1}(t | \bx_i, \gamma_i, \btheta) 
& =  \gamma_i \kappa_1^z \alpha_1^z t^{\alpha_1^z - 1}\expb{\bx_i'\bbeta_1^z} 
\\
\lambda^z_{2}(t | \bx_i, \gamma_i, \btheta) 
& =  \gamma_i \kappa_2^z \alpha_2^z t^{\alpha_2^z - 1}\expb{\bx_i'\bbeta_2^z}
\\
\lambda^z_{3}(t | r_i(z), \bx_i, \gamma_i, \btheta)
& = 
\gamma_i \kappa_3^z \alpha_3^z (t-r_i(z))^{\alpha_3^z - 1} \expb{\bx_i'\bbeta_3^z} & \text{ for } t > r_i(z) 
\end{align*} 
The complete parameter vector for the above model specification is $\btheta = (\balpha, \bbeta, \bkappa, \sigma)$ for $\balpha = (\alpha_1^0, \dots, \alpha_3^1)$, $\bkappa = (\kappa_1^0, \dots, \kappa_3^1)$, and $\bbeta = (\bbeta_1^0, \dots, \bbeta_3^1)'$. For computational convenience we suppose that the independent subject-specific frailties $\gamma_i$ arise from a gamma distribution constrained to have a mean of 1 with unknown variance $\sigma$. This parametric assumption allows the marginal likelihood to be computed analytically, regardless of the specific models used for the baseline hazards. Equation~\ref{eqn:marginallikelihood} gives the likelihood marginalizing over independent gamma-distributed frailties
\begin{align}
\mathcal{L}_{m}
= & \prod_{i=1}^n \bigg[
(1 + \sigma)^{\dri \dti}
[\lambda^{Z_i}_{1}(\yri | \bx_i, \btheta)]^{\dri}
[\lambda^{Z_i}_{2}(\yri | \bx_i, \btheta)]^{\dti(1-\dri)}
[\lambda^{Z_i}_{3}(\yti | \yri, \bx_i, \btheta)]^{\dti\dri} \nonumber \\
& \phantom{ \prod_{i=1}^n} \times 
\bigg( 1 + \sigma \left[ 
\Lambda^{Z_i}_{1}(\yri | \bx_i, \btheta) +
\Lambda^{Z_i}_{2}(\yri | \bx_i, \btheta) +
\Lambda^{Z_i}_{3}(\yti | \yri, \bx_i, \btheta)
\right] \bigg)^{- (1/\sigma + \dri + \dti) } \bigg]
\label{eqn:marginallikelihood}
\end{align}
where $\lambda^z_{j}(t | \bx_i, \btheta) = \lambda^z_{j}(t | \bx_i, \gamma_i = 1, \btheta)$ is a reference level transition hazard for $j\in\{1,2,3\}$. Details of this marginalization can be found in the Web Appendix.

As with any Bayesian procedure, prior distributions must be placed on all unknown parameters. For the frailty variance $\sigma$, we suggest eliciting weakly informative priors from subject matter experts since we encountered convergence problems with dispersed starting values and vague priors (e.g., the $\mathrm{Gamma}(0.7, 0.7)$ prior on the precision $\sigmainv$ suggested by \citet{lee_bayesian_2015}). Priors for other components of $\btheta$ are intended to be weakly informative; details can be found in the Web Appendix. Analyses can be performed with different prior distributions to gauge sensitivity of substantive conclusions to the choice of prior.

\subsection{Estimation algorithm}\label{sec:estimation}
The estimation procedure for the causal quantities can be summarized in four steps: (1) estimating regression model coefficients $\btheta$ using MCMC, (2) sampling latent frailties $\bgamma$ conditional on the posterior of $\btheta$, (3) imputing missing factual and counterfactual outcomes conditional on the posterior of $(\btheta, \bgamma)$, and (4) using imputed potential outcomes to calculate causal estimands of interest.

We obtain posterior samples of $\btheta$ with a modified Hamiltonian Monte Carlo No-U-Turn Sampler in Stan using the marginalized form of the likelihood. Suppose there are $B$ post-warmup MCMC parameter samples $\btheta^{(1)}, \dots, \btheta^{(B)}$. The closed form of $\gamma_i | \btheta$ is a gamma distribution (see Web Appendix), which facilitates sampling from the posterior for $\bgamma$. Using these $B$ posterior samples of $\left(\btheta, \bgamma\right)$, we can draw from the posterior predictive distribution of the full set of potential outcomes $\left(R(Z),\ T(Z),\ R(1-Z),\ T(1-Z) \right)$. Full details of the imputation procedure can be found in the Web Appendix. 

The final step of finite sample causal inference is straightforward once all potential outcomes have been either directly observed or imputed. For a sequence of $K$ time points $t_1, \dots, t_K$ with $t_K \le \max_i(y^R_i)$ dictated by the scientific question, the principal state vector $V(t_k)$ is a deterministic function of $T(0)$ and $T(1)$. For MCMC iteration $b$, denote the principal state for person $i$ at time point $k$ by $V_i(t_k)^{(b)}$. Given $T(0)^{(b)}$ and $T(1)^{(b)}$, let $\left\vert \mathcal{A}_{t_k}^{(b)} \right\vert$ be the number in the always-alive state at $t_k$ (i.e., $\sum_{i=1}^n \bbm{1}\left(V_i(t_k)^{(b)} = AA\right)$). For any $r \in \{t_1, \dots, t_{k}\}$, a posterior draw of the sample time-varying survivor average causal effect is given by
\begin{align*}
TV\text{-}SACE(r, t_k)^{(b)} = &
\left\vert \mathcal{A}_{t_k}^{(b)} \right\vert^{-1} \sum_{i: V_i(t_k)^{(b)} = AA} 
\left[ 
\bbm{1}\left(R_i(1)^{(b)} < r \right) - 
\bbm{1}\left(R_i(0)^{(b)} < r \right)
\right]
\end{align*}

Similarly, the $b^{th}$ posterior draw of the sample restricted mean survivor average effect is
\begin{align*}
RM\text{-}SACE(r, t_k)^{(b)} = &
\left\vert \mathcal{A}_{t_k}^{(b)} \right\vert^{-1} \sum_{i: V_i(t_k)^{(b)} = AA} 
\left[ 
\min\left(R_i(1)^{(b)}, r\right) - 
\min\left(R_i(0)^{(b)}, r\right) 
\right]
\end{align*}
As with any posterior sample, the $B$ draws of $TV\text{-}SACE(r, t_k)^{(b)}$ or $RM\text{-}SACE(r, t_k)^{(b)}$ can be summarized using the means, medians, or quantile-based credible intervals for each ($r, t_k$) pair. Finally, given our reliance on modeling assumptions, some diagnostics are in order.

\subsection{Discrepancy measures for posterior predictive checking}
Certain aspects of the model fit can be assessed by performing posterior predictive checks, which generate replicate data sets $\left(\bX^{rep}, Z^{rep}, Y^{R,rep}, Y^{T,rep}, \delta^{R,rep}, \delta^{T,rep} \right)$. Discrepancy measures are test statistics compared across the observed and replicate data sets, with corresponding $p$-values near 0 or 1 indicating that the assumed data generating process does not explain the data well \citep{gelman_bayesian_2013}. While assumptions such as unconfoundedness remain inherently untestable, these metrics can identify poor model fit to observed features of the data as well as some characteristics of unobserved. 

We propose three discrepancy measures intended to gauge whether causal conclusions are threatened by serious model misspecification. The first set, $T_{KM,z,t}$ for $z \in  \{0,1\}$, are based on the marginal Kaplan-Meier estimates of survival within each treatment group and operate as more traditional goodness of fit tests. These measures deal exclusively with potentially-observable information: survival probabilities $P\left(T(z) > t\right)$ among those patients observed with $Z=z$.  In contrast, another metric ($T_{KS}$) employs Kolmogorov-Smirnov tests to compare the distribution of (imputed) in-sample frailties relative to the distribution implied by the assumed data generation process. Because frailties are inherently latent, $T_{KS}$ can provide evidence of misfit within an unobserved part of the model. A final class of metrics, $T_{AA,t}$, highlight the degree to which frailty misspecification changes estimates of the size of the always-alive stratum. While $T_{KM,0}$ and $T_{KM,1}$ are more traditional model fit assessments, the $T_{KS}$ and $T_{AA}$ discrepancy metrics are different in that their calculation relies on predictions of quantities which are not strictly observed (i.e., involving the dependence between outcomes across treatment conditions).  Details and implementation algorithms for the $T_{KM,z,t}$, $T_{KS}$, and $T_{AA,t}$ metrics are available in the Web Appendix. We provide a complete implementation of steps for parameter sampling, posterior prediction, causal effect estimation, and discrepancy measure calculation through the \texttt{rsemicompstan} R package available on GitHub at github.com/lcomm/rsemicompstan. 

\section{Evaluation of supportive home care effects on mortality and hospital readmission among pancreatic cancer patients}\label{sec:dataapp}
\subsection{Medicare Part A pancreatic cancer readmission data}
We demonstrate our method in an analysis of hospital readmission using a data set of \textcolor{dacolor}{12,091} newly diagnosed pancreatic cancer patients in the United States. The initial sample consisted of \textcolor{dacolor}{17,685} Medicare Part A enrollees in California from 2000 to 2012 who were hospitalized and later discharged with a diagnosis of pancreatic cancer. DWe limited our analysis to the \textcolor{dacolor}{12,091} patients who were healthy enough to be discharged to home (i.e., not hospice or a skilled nursing facility). The baseline $t = 0$ was set to the date of discharge from the index hospitalization during which the cancer was diagnosed. Hospital readmission as a proxy for quality of care usually focuses on a short window after the index hospitalization. To focus on these short-term effects, administrative censoring was applied at 90 days. More information can be found in \citet{lee_bayesian_2015}. 

The scientific question of interest is whether in-home supportive care leads to lower rates of hospital readmission than discharging to home without additional support. Of the \textcolor{dacolor}{12,091} patients discharged to home, \textcolor{dacolor}{3,140 (26\%)} were sent home with supportive care. A major concern was that patients discharged without care would be systematically healthier than those discharged with support, presenting a strong threat of confounding. To reduce the dependence of model-based confounding adjustment, a logistic regression propensity score model for the receipt of home care was constructed using all available baseline covariates: \textcolor{dacolor}{non-White race, age, dichotomized Charleston-Deyo comorbidity score, admission route, and length of stay during index hospitalization}. Estimated propensity scores used to match (without replacement) \textcolor{dacolor}{3,140} of the \textcolor{dacolor}{8,951} patients discharged without care for comparison with those receiving supportive care \citep{ho_matching_2007}. Thus, inference is confined to causal effects among the population represented by those receiving supportive care, i.e., average effects on the ``treated,'' representing the effect of supportive care among those who actually received it.

Hazard regression models included the same covariate set, with all covariates mean centered. Age and length of stay were scaled to have a standard deviation of 1 to facilitate specification of priors for the coefficients. As proposed in Section~\ref{sec:estimation}, we adopted Weibull transition hazards with a semi-Markov specification for the post-readmission hazard of death. \textcolor{dacolor}{Adjustment covariate effects $\bbeta=(\bbeta_1^0,\dots,\bbeta_3^1)'$, the baseline hazard $\bkappa=(\kappa_1^0,\dots,\kappa_3^1)'$, and Weibull shape parameters $\balpha=(\alpha_1^0,\dots,\alpha_3^1)'$ were allowed to freely vary across treatment arm and transition type.} Prior distributions were specified as in Section~\ref{sec:modelspec}.

Posterior draws of $\btheta = (\balpha, \bkappa, \bbeta)$ were obtained from \textcolor{dacolor}{4} chains of \textcolor{dacolor}{4,000} MCMC iterations each, with the first \textcolor{dacolor}{3,000} iterations removed as warmup. Gelman-Rubin potential scale reduction factors $\hat{R}$ and effective sample sizes were calculated for each parameter \citep{gelman_bayesian_2013,carpenter_stan:_2017}. Using the procedure outlined in Section~\ref{sec:estimation}, posterior draws of the frailties and missing potential outcomes were obtained using the \textcolor{dacolor}{4,000} post-warmup samples of $\btheta$.

\subsection{Readmission and mortality results}

All Gelman-Rubin $\hat{R}$ values were below \textcolor{dacolor}{1.01}, indicating good mixing of the chains, and the minimum effective sample size across all parameters was \textcolor{dacolor}{2,455}. 

\begin{figure}
\centering
\includegraphics[width=0.98\textwidth]{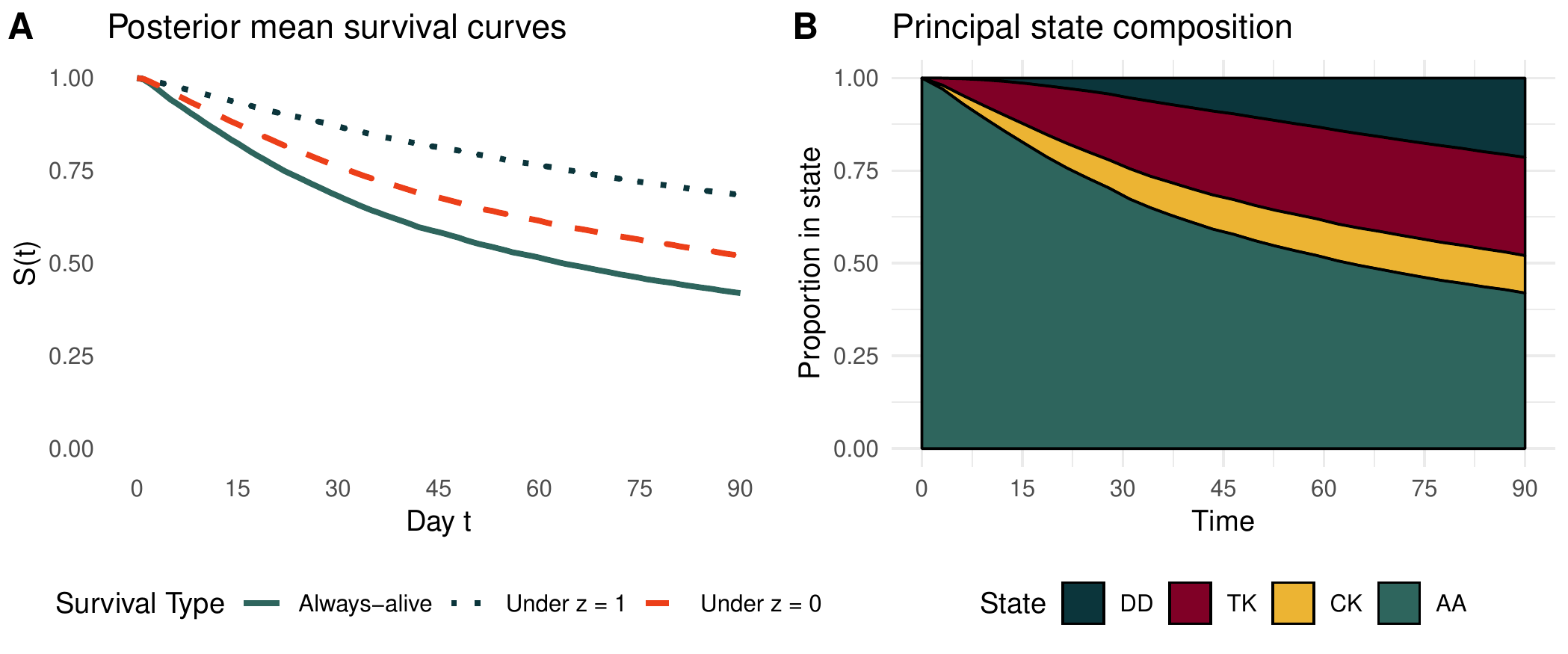}
\caption{Posterior mean survival curves among newly diagnosed pancreatic cancer patients discharged home, with supportive care ($z=1$) and without ($z=0$), with the corresponding implications for always-alive principal stratum size (A) and posterior mean population composition of always-alive ($AA$), treatment-killed ($TK$), control-killed ($CK$), and doubly dead ($DD$)  principal states (B)}\label{fig:dastatepair}
\end{figure}

Part A of Figure~\ref{fig:dastatepair} shows the posterior mean survival curves for each treatment group and their implications for posterior mean size of the always-survivor subpopulation. The fastest-declining survival curve, shown in green, is the in-sample average of time to first potential death (i.e., $\min\left(T_i(0), T_i(1)\right)$); the ``survival'' $S(t)$ equals $P\left(V(t) = AA\right)$. The other two curves in Part A show the mean $S(t)$ for the counterfactual survival probabilities if everyone in the (matched) sample had been treated \textcolor{dacolor}{with extra care} ($z=1$, in navy) or \textcolor{dacolor}{discharged home without extra care} ($z=0$, in orange). Based on these covariate-adjusted survival curves, the treatment of receiving additional support at home leads to \textcolor{dacolor}{reduced} lifespan across the 90 days, i.e., $P\left(T(0) > t\right) > P\left(T(1) > t\right)$ for $t < 90$. For all curves, uncertainty increases with time because of the decreasing number of subjects used to estimate survival. Relative to the $T(z)$ survival curves, there is additional uncertainty in the $P\left(V(t) = AA\right)$ estimates due to uncertainty in $\sigma$ and $\bgamma$.

Part B of Figure~\ref{fig:dastatepair} shows the posterior mean proportion of the population in each principal state $\{AA, TK, CK, DD\}$ over time. For small $t$, nearly the entire population is in the $AA$ state because few deaths are observed or imputed under $z \in \{0,1\}$. With time, more deaths accumulate among patients discharged home \textcolor{dacolor}{with support}, leading to a \textcolor{dacolor}{greater} proportion of the population in the $TK$ state than the $CK$ state. The population fractions in $TK$ and $CK$ \textcolor{dacolor}{stay relatively constant after approximately day 45, suggesting that most patients who would die \emph{only} if discharged to one of the conditions will do so relatively early in the 90-day time frame}. The overall effect is that depletion of the always-alive principal stratum occurs more during the early part of the 90-day window.

\subsubsection{Population-level causal effects}
\begin{figure}
\centering
\includegraphics[width=0.95\textwidth]{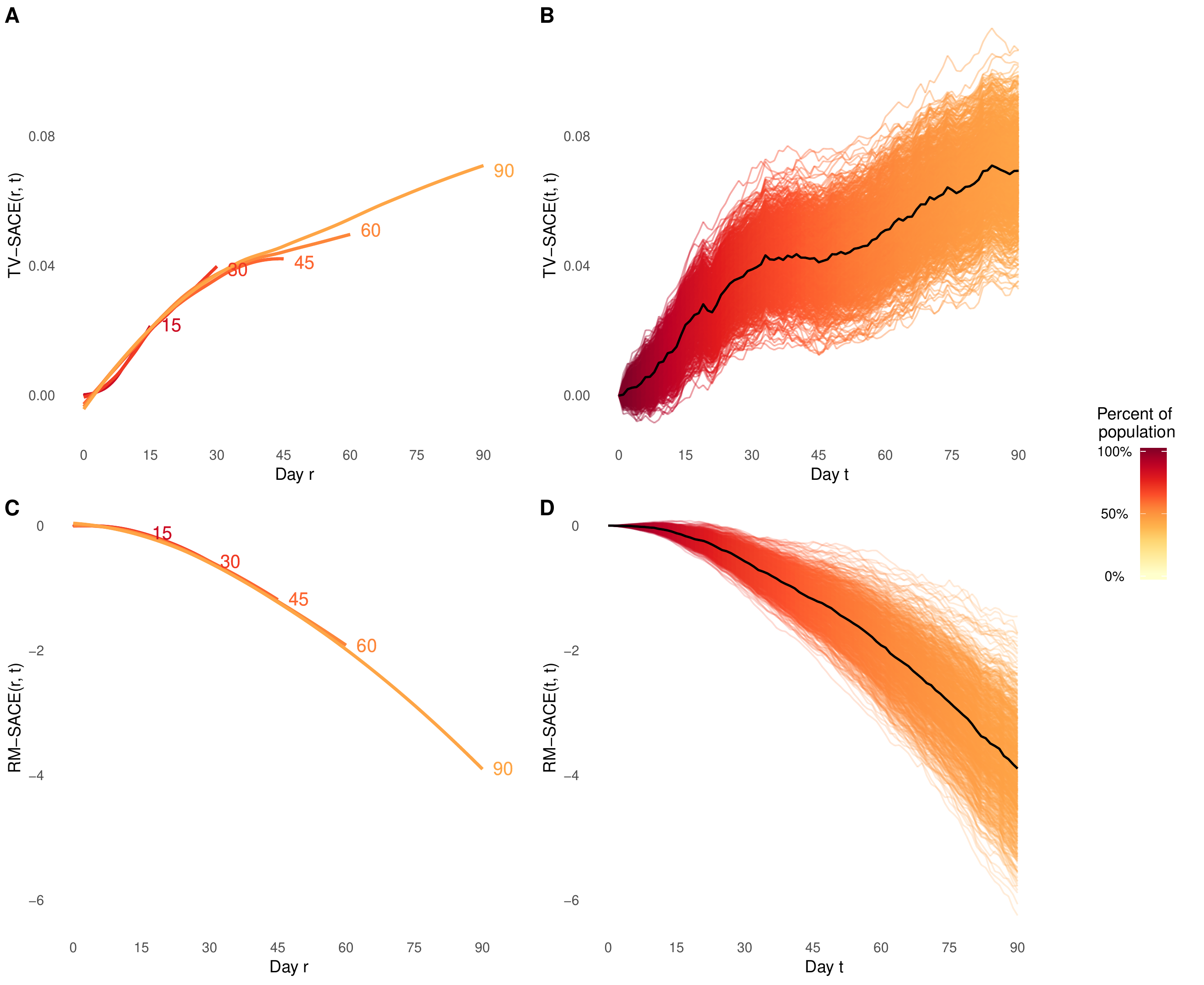}
\caption{Estimated time-varying ($TV\text{-}SACE$) and restricted mean ($RM\text{-}SACE$) survivor average causal effects of home care (vs. no additional care at home) on the cumulative incidence of hospital readmission among 6,280 newly diagnosed late-stage pancreatic cancer patients} \label{fig:dasacequad}
\end{figure}

Applying principal stratification to semicompeting risks data allows us to characterize treatment effects among subgroups defined by always-survivorship through various $t$ (i.e., $\mathcal{A}_t$ for various $t$). Part A of Figure~\ref{fig:dasacequad} shows posterior means for $TV\text{-}SACE(r,t)$ for 5 always-survivor cohorts $\mathcal{A}_t$: $t \in \{ 15,30,45,60,90\}$. In all cohorts, \textcolor{dacolor}{support} leads to \textcolor{dacolor}{greater} incidence of hospital readmission. In the first days after discharge from the index hospitalization, the healthier, longer-surviving cohorts like $\mathcal{A}_{90}$ have treatment effects on readmission rates which are \textcolor{dacolor}{similar to} cohorts with less stringent survivorship requirements (e.g., $\mathcal{A}_{15}$). However, effects among the longer-surviving cohorts \textcolor{dacolor}{begin to level off over time}. This may point to a heterogeneity in reasons requiring a readmission; that is, readmissions occurring in the first week or so after diagnosis may be caused by a different mixture of proximate causes than the admissions during the rest of the 90 days. Additional contact with medical personnel at home may also speed the detection of the early complications warranting readmission. However, given the poorer survival of the supported group, this could also be the result of residual uncontrolled confounding by indication if, after adjusting for $\bX$ and the propensity score preprocessing, those receiving supportive care remain at systematically higher risk for readmission.

From a policy perspective, we may be interested in the consistency of snapshots effect across time. If the estimated effects on readmission vary dramatically depending on $t$, then policymakers must be more careful when synthesizing evidence across studies offering snapshots from different $t$. Plot B of Figure~\ref{fig:dasacequad} shows estimated curves of $Q(t) = TV\text{-}SACE(t,t)$ across $t$, with each of the 500 lines derived from a representative posterior draw of $\btheta$. The color of the lines at each $t$ gives the proportion of the study population in the always-alive state at $t$ according to that set of posterior predictive potential outcome samples. The shape suggests that, for the cumulative incidence scale, there is no natural time point for evaluating the causal effect of discharge support on hospital readmission because $Q(t)$ never completely levels off. However, the direction of the effect (i.e., higher cumulative incidence in the group discharged with care) is largely consistent over time.

Like the time-varying survivor average causal effect, the restricted mean effects also suggest that \textcolor{dacolor}{being discharged home with support} \textcolor{dacolor}{increases} readmissions. Part C of Figure~\ref{fig:dasacequad} shows the within-cohort accumulation of readmission-free days attributable to being \textcolor{dacolor}{discharged with support}. Because the accumulation is \textcolor{dacolor}{negative}, this finding is consistent with faster and ultimately \textcolor{dacolor}{greater cumulative incidence of readmission among the treated (i.e., supported)} group. In part due to the natural ceiling of $t$ in the definition of the restricted mean, the estimated snapshot function $M(t)=RM\text{-}SACE(t,t)$ in Part D of Figure~\ref{fig:dasacequad} steadily grows in magnitude over the course of the 90 days.

\subsubsection{Implications for individual-level decisionmaking}
The posterior distribution for frailty variance $\sigma$ shows substantial remaining variability in prognosis that is not explained by the covariates included in the models, with a mean of \textcolor{dacolor}{1.44 (95\% CI: 1.25, 1.65)}. To put this estimate into perspective, \textcolor{dacolor}{$\sigma = 1.44$} corresponds to patients in the 90$^{th}$ percentile of the latent frailty experiencing event hazards that are \textcolor{dacolor}{55} times the hazards for comparable patients in the 10$^{th}$ percentile. Relative to the variation in prognoses explained by predictive covariates, large values for $\sigma$ pose additional difficulties for tailored decisionmaking. Nevertheless, covariate-specific posterior predictions may be used to differentiate treatment recommendations.

In contrast to policymaking motivated by population effects, decisionmakers choosing for a single individual may be interested in the extent to which that individual's readmission prospects depend on treatment, as well as treatment's impact on survival. Table~\ref{tab:pps} gives examples of tailored prognoses for a two selected covariate patterns. If more information is known about the underlying health state than the covariates used in the initial analysis -- say, the individual can infer they are healthier than average patients with similar observed covariates -- these prognoses can be further personalized.

\begin{table}
\caption{\label{tab:pps}Posterior predictive  means for principal  state probabilities and principal stratum causal effects for new patients of two covariate patterns}
\centering

\tiny \begin{threeparttable}[t] \begin{tabular}{llcccccc}\toprule
               \multicolumn{1}{c}{} & \multicolumn{1}{c}{} &  & 
               \multicolumn{3}{c}{\begin{tabular}[c]{@{}c@{}}Principal State \\ 
               Probabilities at $t$\tnote{1}\end{tabular}} & 
               \multicolumn{2}{c}{\begin{tabular}[c]{@{}c@{}}If always-alive, \\
               causal effect of being \\ 
               discharged to home with support \\ (vs. without) \end{tabular}} 
               \\ \midrule
Patient characteristics & Latent health\tnote{2} & Day $t$ & AA & CK & TK & \begin{tabular}[c]{@{}c@{}}Difference in readmission \\ incidence by $t$ \end{tabular} & \begin{tabular}[c]{@{}c@{}}Additional readmission-free \\ days accumulated by $t$ \end{tabular}\\
\midrule
\multirow[t]{6}{*}{\begin{tabular}[t]{@{}l@{}}Nonwhite male aged 85, \\ average comorbidity score \\ and hospital length of stay\end{tabular}} & \multirow[t]{2}{*}{Frail} & 30 & 0.181 & 0.140 & 0.383 & -0.173 & 5.444\\
 &  & 90 & 0.003 & 0.013 & 0.159 & -0.040 & 14.170\\
 & \multirow[t]{2}{*}{Average} & 30 & 0.560 & 0.122 & 0.261 & -0.128 & 2.821\\
 &  & 90 & 0.107 & 0.106 & 0.394 & -0.101 & 12.116\\
 & \multirow[t]{2}{*}{Healthy} & 30 & 0.979 & 0.007 & 0.014 & -0.009 & 0.152\\

 &  & 90 & 0.916 & 0.026 & 0.056 & -0.018 & 1.000\\
\multirow[t]{6}{*}{\begin{tabular}[t]{@{}l@{}}White female aged 65, \\ average comorbidity score \\ and hospital length of stay\end{tabular}} & \multirow[t]{2}{*}{Frail} & 30 & 0.520 & 0.157 & 0.248 & -0.024 & 1.387\\
 &  & 90 & 0.110 & 0.157 & 0.301 & 0.000 & 1.793\\
 & \multirow[t]{2}{*}{Average} & 30 & 0.821 & 0.063 & 0.107 & -0.030 & 0.966\\
 &  & 90 & 0.449 & 0.162 & 0.286 & 0.002 & 1.470\\

 & \multirow[t]{2}{*}{Healthy} & 30 & 0.995 & 0.001 & 0.004 & -0.003 & 0.076\\
 &  & 90 & 0.979 & 0.006 & 0.015 & 0.000 & 0.200\\
\bottomrule\end{tabular}\begin{tablenotes}
                \item[1] Always-alive (AA), dead only under control (CK), and dead only under treatment (TK)
                \item[2] Frail and healthy correspond to the $90^{th}$ and
               $10^{th}$ percentiles of
               $\gamma$, while average health corresponds to $\gamma = 1$
               \end{tablenotes}
               \end{threeparttable}

\end{table}

As expected, the posterior predictive state probabilities show that -- for comparable levels of underlying frailness -- a younger White woman is much more likely to be in the always-alive state at 90 days than an older non-White man with the same comorbidity score and duration of index hospitalization. However, the magnitude of this survival advantage varies greatly. For an individuals in the $10^{th}$ or $90^{th}$ percentile of latent health (i.e., the $90^{th}$ or $10^{th}$ percentile for $\gamma$), the difference in the probability of being always-alive at $t=90$ is approximately \textcolor{dacolor}{0.06} to \textcolor{dacolor}{0.11}; for individuals of average frailty, the difference is more pronounced at \textcolor{dacolor}{0.342 (0.449 vs. 0.107)}. We can also conclude that frail patients of either covariate pattern are unlikely to be in the always-alive state at 90 days. Together, these findings highlight the large degree to which an individual may be able to tailor their decisionmaking based on additional information.

\subsubsection{Assessment of model fit with discrepancy measures}
\begin{figure}
\centering
\includegraphics[width=0.95\textwidth]{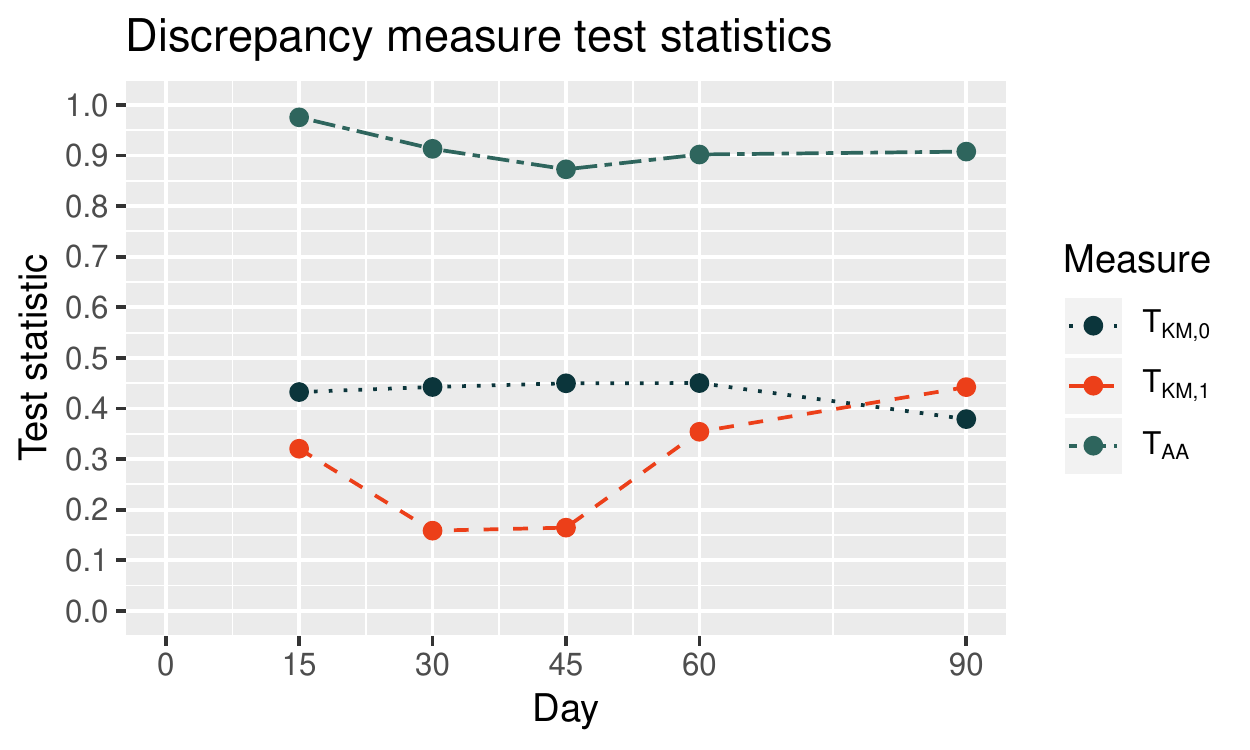}
\caption{Posterior predictive $p$-values for three discrepancy measures assessing model fit with respect to proportion always-alive ($T_{AA}$), and marginal survival under treatment ($T_{KM,1}$) and control ($T_{KM,0}$)} \label{fig:dadiscrep}
\end{figure}

Figure~\ref{fig:dadiscrep} shows the posterior predictive $p$-values from the metrics for marginal survival under treatment ($T_{KM,1,t}$) and control ($T_{KM,0,t}$). While the $T_{KM,1}$ and $T_{KM,0}$ $p$-values stay far from either extreme-- indicating adequate fit-- the $T_{KS}$ $p$-value (\textcolor{dacolor}{$<0.001$}, not shown in figure) suggests potential misspecification of the underlying frailty distribution. As can be seen in Figure~\ref{fig:dadiscrep} with $T_{AA,t}$ values near \textcolor{dacolor}{0.9}, this form of misspecification has a moderate impact on the size of the always-alive population. In particular, the observed sample appears to have an unusually \textcolor{dacolor}{large} proportion in the always-alive state. 

\section{Conclusions}
In this paper we have proposed a general approach to principal stratification where the strata are defined by potential times to a truncating event. From a decisionmaking perspective, this stratification is a natural one because it groups units according to their time horizon for comparing quality of life. To quantify those differences, we formulated two new causal estimands, the $TV\text{-}SACE$ and $RM\text{-}SACE$, for contrasting nonterminal time-to-event outcomes that are truncated by death. We then described a Bayesian model-based estimation procedure that builds upon existing strategies for semicompeting risks models. Our innovative factorization scheme facilitates connections to existing illness-death models, putting a sharper causal focus on this literature and clarifying how such models can be adapted to yield causally interpretable quantities. 

The methods outlined here have several limitations that warrant discussion. First, in our implementation, the latent trait is assumed to be gamma-distributed, and the effect of the frailty is constrained to be identical across all hazard types and treatment arms. Both assumptions were made largely for computational convenience because they allow parameter sampling using the marginal likelihood. Other parametric distributions could be assumed for the latent trait (e.g., log-normal), and shared frailty models have previously incorporated transition-specific coefficients for the log-frailty \citep{liu_shared_2004}. These adaptations do not result in analytically tractable marginal likelihoods, although numerical integration can be used. In practice, we found MCMC performance using the unmarginalized likelihood to be inconsistent, slow, and prone to divergent NUTS transitions. Second, the specific parametric Weibull hazard models may not be appropriate for all scenarios. More flexible baseline hazard specifications could be achieved with splines \citep{royston_flexible_2002} or Bayesian nonparametrics \citep{lee_bayesian_2015}, although posterior prediction would become more difficult due to problems extrapolating beyond the observed time scale.

Notwithstanding these limitations, this work offers a new, causally informed approach to the analysis of semicompeting risks data. Illness-death models pose two challenges for causal inference on the nonterminal outcome: (1) the use of hazard-based estimation strategies, which implicitly condition on the post-treatment outcome of survival, and (2) handling truncation when the treatments also affect survival. By formulating causal estimands using potential outcomes notation, we separate the model estimation from the choice of causal estimand. Our method also indirectly addresses decisionmakers' need to balance non-mortality considerations with treatment impacts on survival; this is achieved by quantifying, for every time point, the relative size of the population for whom quality of life contrasts are relevant. The use of posterior predictive sampling to estimate the effects allows for the innovative density factorization which connects to an existing semicompeting risks approach. Analogous factorizations may prove useful for truncated outcomes which are not time-to-event. Lastly, because we operate in the Bayesian framework, we properly account for uncertainty due to partial identifiability of the causal effects. 

Future work on these and other principal stratification models for quality of life outcomes in high-mortality settings may be extended to incorporate utility functions within a formal decision-theoretic framework. These methods open up more possibilities for causally valid research on non-mortality outcomes related to quality of life among high-mortality patient populations. In turn, this provides evidence that is more directly useful to the individuals and policymakers who must balance considerations of survival and quality of life.





\section*{Acknowledgements}
Support for this work was provided by NIH grants T32CA009337 and T32ES007142 (LC), R01CA181360 (SH), and R01ES026217 and R01GM111339 (CZ), as well as EPA grant RD835872 (CZ). FM received support from Dipartimenti Eccellenti 2018-2022 ministerial funds. LC was also supported by a Rose Traveling Fellowship. We thank Alessandra Mattei for helpful discussions about discrepancy measures.


\newpage
\section*{Supplementary Materials}
\appendix

\section{Additional details on the prior specification}\label{sec:appendixpriordetails}
For binary covariates and continuous variables rescaled to have unit variance, hazard ratios are unlikely to exceed 5; therefore we set $\pi(\bbeta_j^z)$ to be $\mathcal{N}(0, 2.5^2)$ for $j=1,2,3$ and $z=0,1$. With mean-centered covariates and $\alpha_j^z = 1$, the baseline hazard $\kappa_j^z$ corresponds to the hazard experienced by those at the sample mean covariate values. Thus, reasonable priors for the log-baseline hazards are $\mathcal{N}(\log(E_j/PT_j), \left(\log(100)/2 \right)^2)$, where $E_j$ is the number of observed events and $PT_j$ is the total at-risk person-time for transition $j$, pooling across treatment arms. For exponential hazards, this prior asserts that the true hazard experienced at the sample mean value has only $\approx$ 5\% probability of being more than two orders of magnitude away from the crude (pooled) event rate. The data-driven prior specification for the log-baseline hazards makes the model invariant to the time scale of the data (i.e., days vs. years). (An alternative approach would be to rescale all times so that the mean event times were $\approx 1$.) Lastly, the Weibull shape parameters $(\alpha_1^{0},\dots,\alpha_3^{1})$ are given $\mathrm{LogNormal}(0, 2^2)$ priors to express moderate belief that any changes in the hazards occur slowly rather than quickly decaying or exploding. Assuming variation independence of the different parameter blocks $\bbeta$, $\balpha$, and $\bkappa$, we can construct a prior as
\begin{equation}
\pi(\btheta) = \pi(\bbeta) \pi(\balpha) \pi(\bkappa)
\end{equation}

\section{Marginalization of conditional likelihood $\mathcal{L}_c$ over frailties $\bgamma$}

Let $\lambda_j(t|\cdot)$ be the instantaneous hazards and $\Lambda_j(t|\cdot)$ be the cumulative hazards for transition $j \in \{ 1,2,3\}$. Then the conditional likelihood contribution of individual $i$ is
\begin{align}
\mathcal{L}_{c, i}
=
&
[ \lambda^{Z_i}_{1}(\yri | \bx_i, \gamma_i, \btheta) ]^{\dri}
[\lambda^{Z_i}_{2}(\yri | \bx_i, \gamma_i, \btheta)]^{\dti(1-\dri)}
[\lambda^{Z_i}_{3}(\yti | \yri, \bx_i, \gamma_i, \btheta)]^{\dti\dri}
\nonumber \\
& \times 
\expb{ -
\Lambda^{Z_i}_{1}(\yri | \bx_i, \gamma_i, \btheta) -
\Lambda^{Z_i}_{2}(\yri | \bx_i, \gamma_i, \btheta) - 
\Lambda^{Z_i}_{3}(\yti | \yri, \bx_i, \gamma_i, \btheta)} 
\end{align}

 with $f(\bgamma|\sigma) = \prod_{i=1}^n \left[\sigma^{1/\sigma} \Gamma(\sigmainv) \right]^{-1}
\gamma_i^{1/\sigma - 1}\expb{- \gamma_i (1/\sigma) }$, as in the main text. The marginal likelihood across all observations can be written as 
\begin{align}
\mathcal{L}_m = & \prod_{i=1}^n \left( \int_{0}^\infty \mathcal{L}_{c,i} f(\gamma_i | \sigma) d\gamma_i \right) \equiv \prod_{i=1}^n \mathcal{L}_{m,i}
\end{align}
This was first stated in \citet{xu_statistical_2010}, but we provide a proof here.

First, we define some shorthand notation to suppress indexing that is unnecessary for this proof:
\begin{align*}
h_1 = & \lambda^{Z_i}_{1}(\yri | \bx_i, \gamma_i, \btheta) \\
h_2 = & \lambda^{Z_i}_{2}(\yri | \bx_i, \gamma_i, \btheta) \\
h_3 = & \lambda^{Z_i}_{3}(\yti | \yri, \bx_i, \gamma_i, \btheta) \\
H_1 = & \Lambda^{Z_i}_{1}(\yri | \bx_i, \gamma_i, \btheta) \\
H_2 = & \Lambda^{Z_i}_{2}(\yri | \bx_i, \gamma_i, \btheta) \\
H_3 = & \Lambda^{Z_i}_{3}(\yti | \yri, \bx_i, \gamma_i, \btheta) \\
s = & \sigma \\
d_e = & \delta_i^E \text{ for }E \in \{R, T \} \\
g = & \gamma_i
\end{align*}
and note a general property of the gamma function that for $k \in \mathbb{R}^+$
\begin{align}
\Gamma(k + 1) = k \Gamma(k) \label{eq:gammaproperty}
\end{align}

\begin{align*}
\mathcal{L}_{m,i}
= & 
\int_{0}^{\infty} \frac{g^{1/s-1}}{s^{1/s}\Gamma(1/s)} h_1^\dr h_2^{\dt(1-\dr)} h_3^{\dr\dt} \expb{- H_1 - H_2 - H_3 - g/s} dg \\
= & \frac{s^{-1/s}{\Gamma(1/s + \dr + \dt)} h_1^{\dr} h_2^{\dt(1-\dr)} h_3^{\dr\dt} \left(1/s + H_1 + H_2 + H_3\right)^{-1/s - \dr - \dt}}{\Gamma(1/s)} \\
= & \frac{\Gamma(1/s + \dr + \dt)}{\Gamma(1/s)} \times 
h_1^{\dr} h_2^{\dt(1-\dr)} h_3^{\dr\dt} \times
s^{(\dr + \dt)} \underbrace{s^{-1/s - {\dr} - {\dt}}\left(1/s + {H_1} + {H_2} + {H_3} \right)^{-1/s - {\dr} - {\dt}}}_{\left( 1 + s({H_1} + {H_2} + {H_3}) \right)^{-1/s - {\dr} - {\dt}}} \\
= & h_1^{\dr} h_2^{\dt(1-\dr)} h_3^{\dr\dt} \times
\underbrace{(1 + s)^{\dr \dt}}_{\left(s^{\dr+\dt}\right)\frac{\Gamma(1/s + \dr + \dt)}{\Gamma(1/s)} \text{ for the relevant }(\protect{\dr},\protect{\dt})} \times 
\left( 1 + s(H_1 + H_2 + H_3) \right)^{-1/s - \protect{\dr} - \protect{\dt}}
\end{align*}

To see the final line, consider the 4 possible values taken on by the binary indicators $(\dr, \dt)$:

\textbf{Case 1: $(\dr, \dt) = (0,0)$}
\begin{align*}
1 = (1 + s)^0 = 
& s^{(0 + 0)} \frac{\Gamma(1/s)}{\Gamma(1/s)} = 1
\end{align*}

\textbf{Cases 2 and 3: $(\dr, \dt) = (0,1)$ and $(1,0)$}
\begin{align*}
1 = (1 + s)^0 = 
& \frac{s^1 \Gamma(1/s + 1)}{\Gamma(1/s)} = 1
\end{align*}
where the rightmost equality is true by Equation~\ref{eq:gammaproperty}.

\textbf{Case 4: $(\dr, \dt) = (1,1)$}
\begin{align*}
s^{1 + 1} \frac{\Gamma(1/s + 1 + 1)}{\Gamma(1/s)} 
= & s (1/s)^{-1} \underbrace{\frac{\Gamma(1/s + 2)}{\Gamma(1/s + 1)}}_{1/s + 1 \text{ by \ref{eq:gammaproperty}}} \frac{\Gamma(1/s + 1)}{\Gamma(1/s)} \\
= & s (1/s + 1) \times \underbrace{\frac{\Gamma(1/s + 1)}{(1/s) \Gamma(1/s)}}_{\text{1 by \ref{eq:gammaproperty}}} \\
= & s + 1\\
\end{align*}

This proves the marginal likelihood has the form stated in the main text:
\begin{align*}
\mathcal{L}_{m}
= & \prod_{i=1}^n \bigg[
(1 + \sigma)^{\dri \dti}
[\lambda^{Z_i}_{1}(\yri | \bx_i, \btheta)]^{\dri}
[\lambda^{Z_i}_{2}(\yri | \bx_i, \btheta)]^{\dti(1-\dri)}
[\lambda^{Z_i}_{3}(\yti | \yri, \bx_i, \btheta)]^{\dti\dri} \nonumber \\
& \phantom{ \prod_{i=1}^n} \times 
\bigg( 1 + \sigma \left[ 
\Lambda^{Z_i}_{1}(\yri | \bx_i, \btheta) +
\Lambda^{Z_i}_{2}(\yri | \bx_i, \btheta) +
\Lambda^{Z_i}_{3}(\yti | \yri, \bx_i, \btheta)
\right] \bigg)^{- (1/\sigma + \dri + \dti) } \bigg]
\end{align*}

\section{Log-likelihood contributions by observed data pattern}

The log-likelihood marginalized over the frailties is
\[
\ell_{mi} 
= 
\dri \dti \bigg( \log \left(1 + \sigma \right)
+ \log (\lambda_3) \bigg) + 
\dri \log (\lambda_1) + \dti(1-\dri) \log(\lambda_2)
-(1/\sigma + \dri + \dti) 
\log (1 + B)
\]
with $B = \sigma(\Lambda_1 + \Lambda_2 + \Lambda_3)$. This is the likelihood that gets added to the \texttt{target} function within Stan.

The marginal likelihood $\calL_m$ in the main text corresponds to 4 types of marginal likelihood and log-likelihood contributions: (1) neither event occurrence, (2) nonterminal occurrence only, (3) terminal occurrence only, and (4) both event occurrence.

\begin{enumerate}
\item Observe neither event ($\dri = \dti = 0$)
\begin{align*}
\calL_{mi} = & \bigg( 1 + \sigma \left[ 
\Lambda^{Z_i}_{1}(\yri | \bx_i, \btheta) +
\Lambda^{Z_i}_{2}(\yri | \bx_i, \btheta)
\right] \bigg)^{- (1/\sigma) }
\\
\ell_{mi} = & -(1/\sigma) \log \left( 1 + \sigma \left[ 
\Lambda^{Z_i}_{1}(\yri | \bx_i, \btheta) +
\Lambda^{Z_i}_{2}(\yri | \bx_i, \btheta) \right] \right) 
\\
= & 
-(1/\sigma) 
\log \left( 1 + \sigma \left[ 
\kappa_1^{Z_i}e^{\bx_i'\bbeta^z_1}  (\yri)^{\alpha_1^{Z_i}} +
\kappa_2^{Z_i} e^{\bx_i'\bbeta^z_2} (\yri)^{\alpha_2^{Z_i}}  
\right] \right) 
\end{align*}
\item Observe only nonterminal ($\dri = 1, \dti = 0$)
\begin{align*}
\calL_{mi} = & 
\lambda^{Z_i}_{1}(\yri | \bx_i, \btheta)
\bigg( 1 + \sigma \left[ 
\Lambda^{Z_i}_{1}(\yri | \bx_i, \btheta) +
\Lambda^{Z_i}_{2}(\yri | \bx_i, \btheta) +
\Lambda^{Z_i}_{3}(\yti | \yri, \bx_i, \btheta)
\right] \bigg)^{- (1/\sigma + 1) }
\\
\ell_{mi} = &
\log\left(\lambda^{Z_i}_{1}(\yri | \bx_i, \btheta)\right)
+ \nonumber \\
& -(1/\sigma + 1)
\log\left( 1 + \sigma \left[ 
\Lambda^{Z_i}_{1}(\yri | \bx_i, \btheta) +
\Lambda^{Z_i}_{2}(\yri | \bx_i, \btheta) +
\Lambda^{Z_i}_{3}(\yti | \yri, \bx_i, \btheta) \right] \right) 
\\
= & 
\log \left( \kappa_1^{Z_i} \alpha_1^{Z_i} e^{\bx_i'\bbeta^z_2} (\yri)^{\alpha_1^{Z_i}}\right) + \nonumber \\
& -(1/\sigma + 1)
\log\left( 1 + \sigma \left[ 
\kappa_1^{Z_i}e^{\bx_i'\bbeta^{Z_i}_1}  (\yri)^{\alpha_1^{Z_i}} +
\kappa_2^{Z_i} e^{\bx_i'\bbeta^{Z_i}_2} (\yri)^{\alpha_2^{Z_i}} +
\kappa_3^{Z_i}e^{\bx_i'\bbeta^{Z_i}_3}  (\yti - \yri)^{\alpha_3^{Z_i}}
\right] \right)
\end{align*}
\item Observe only terminal ($\dri = 0, \dti = 1$)
\begin{align*}
\calL_{mi} = & \lambda^{Z_i}_{2}(\yri | \bx_i, \btheta)
\bigg( 1 + \sigma \left[ 
\Lambda^{Z_i}_{1}(\yri | \bx_i, \btheta) +
\Lambda^{Z_i}_{2}(\yri | \bx_i, \btheta)
\right] \bigg)^{- (1/\sigma + 1) } 
\\
\ell_{mi} = &
\log\left( \lambda^{Z_i}_{2}(\yri | \bx_i, \btheta) \right) + \nonumber\\
& -(1/\sigma + 1)
\log\left( 1 + \sigma \left[ 
\Lambda^{Z_i}_{1}(\yri | \bx_i, \btheta) +
\Lambda^{Z_i}_{2}(\yri | \bx_i, \btheta)
\right] \right) 
\\
= & 
\log \left( \kappa_2^{Z_i} \alpha_2^{Z_i} e^{\bx_i'\bbeta^{Z_i}_2} (\yri)^{\alpha_2^{Z_i}} \right) + \nonumber \\
& -(1/\sigma + 1)
\log\left( 1 + \sigma \left[ 
\kappa_1^{Z_i}e^{\bx_i'\bbeta^{Z_i}_1}  (\yri)^{\alpha_1^{Z_i}} +
\kappa_2^{Z_i} e^{\bx_i'\bbeta^{Z_i}_2} (\yri)^{\alpha_2^{Z_i}}
\right] \right)
\end{align*}
\item Observed both events ($\dri = \dti = 1$)
\begin{align*}
\calL_{mi} = &
(1 + \sigma)
\lambda^{Z_i}_{1}(\yri | \bx_i, \btheta)
\lambda^{Z_i}_{3}(\yti | \yri, \bx_i, \btheta) \times \nonumber \\
& \bigg( 1 + \sigma \left[ 
\Lambda^{Z_i}_{1}(\yri | \bx_i, \btheta) +
\Lambda^{Z_i}_{2}(\yri | \bx_i, \btheta) +
\Lambda^{Z_i}_{3}(\yti | \yri, \bx_i, \btheta)
\right] \bigg)^{- (1/\sigma + \dri + \dti) } 
\\
\ell_{mi} = &
\log \left(1 + \sigma \right) + 
\log\left( \lambda^{Z_i}_{1}(\yri | \bx_i, \btheta) \right) + 
\log\left(\lambda^{Z_i}_{3}(\yti | \yri, \bx_i, \btheta)\right) + \nonumber \\
& -(1/\sigma + 2)
\log \left( 1 + \sigma \left[ 
\Lambda^{Z_i}_{1}(\yri | \bx_i, \btheta) +
\Lambda^{Z_i}_{2}(\yri | \bx_i, \btheta) +
\Lambda^{Z_i}_{3}(\yti | \yri, \bx_i, \btheta)
\right] \right)
\\
= & 
\log \left(1 + \sigma \right) + 
\log \left( \kappa_1^{Z_i} \alpha_1^{Z_i} e^{\bx_i'\bbeta^{Z_i}_1} (\yri)^{\alpha_1^{Z_i}} \right) +
\log \left( \kappa_3^{Z_i} \alpha_3^{Z_i} e^{\bx_i'\bbeta^{Z_i}_3} (\yti - \yri)^{\alpha_3^{Z_i}} \right) + \nonumber \\
& -(1/\sigma + 2)
\log \left( 1 + \sigma \left[ 
\kappa_1^{Z_i}e^{\bx_i'\bbeta^{Z_i}_1} (\yri)^{\alpha_1^{Z_i}} + 
\kappa_2^{Z_i} e^{\bx_i'\bbeta^{Z_i}_2} (\yri)^{\alpha_2^{Z_i}} + 
\kappa_3^{Z_i} e^{\bx_i'\bbeta^{Z_i}_3} (\yti - \yri)^{\alpha_3^{Z_i}}
\right] \right)
\end{align*}
\end{enumerate}

\section{Frailty marginalization and posterior predictive imputation}
\subsection{Full conditional form of frailties}\label{sec:fullcondfrailties}
Omitting terms which do not depend on $\gi$, the conditional likelihood as a function of $\gi$ is
\begin{equation*}
\mathcal{L}_{c}
\propto 
\gi^{\dri + \dti} \expb{-\gi \left[ 
\Lambda_{1}(\yri | \bx_i, \btheta) + 
\Lambda_{2}(\yri | \bx_i, \btheta) + 
\Lambda_{3}(\yti | \yri, \bx_i, \btheta) + 
\right]}
\end{equation*} 
This demonstrates that the posterior distribution of $\gi$, conditional on $\btheta$, only depends on the data through $O_i$. The only other place $\gi$ appears in the posterior is in $f(\gi|\sigma)$, which has kernel $\gi^{1/\sigma - 1}\exp\{-\gi (1/\sigma)\}$. Thus, the full conditional distribution for $\gi$ is
\begin{equation*}  
\pi(\gi | \cdot) \propto 
\gi^{1/\sigma + \dri + \dti - 1} \expb{-\gi \left[ 
 1/\sigma + 
\Lambda_{1}(\yri | \bx_i, \btheta) + 
\Lambda_{2}(\yri | \bx_i, \btheta) + 
\Lambda_{3}(\yti | \yri, \bx_i, \btheta)
\right]}
\end{equation*}
which can be recognized as the kernel of a $\mathrm{Gamma}(a_1,a_2)$ with $a_1 = 1/\sigma + \dri + \dti$ and $a_2 = 1/\sigma + \Lambda_{1}(\yri | \bx_i, \btheta) + 
\Lambda_{2}(\yri | \bx_i, \btheta) + 
\Lambda_{3}(\yti | \yri, \bx_i, \btheta)$.

\subsection{Frailty imputation}
For each $i=1,\dots,n$ and $b=1,\dots,B$, sample $\gib$ as
\begin{equation*}
\gib | \btheta^{(b)}
\sim
\mathrm{Gamma}\left(
1/\sigma + \dri + \dti, \ 
1/\sigma + \Lambda_{1}(\yri | \bx_i, \bthetab) + \Lambda_{2}(\yri | \bx_i, \bthetab) + \Lambda_{3}(\yti | \yri, \bx_i, \bthetab) 
\right)
\end{equation*} 

\subsection{Sampling from the posterior predictive distribution}\label{sec:imppo}
\subsubsection{Imputation of censored outcomes}\label{sec:impcens}
Censoring is the cause of missing outcome data in the factual treatment arm. In the presence of censoring for individual $i$, there is only partial information on one or both of $\left(R_i(Z_i), T_i(Z_i)\right)$. Given censoring time $C_i$ and draw $b$ of the posterior parameter and frailty vectors $(\bthetab, \gib)$, we can impute $R_i^{(b)}(Z_i)$ or $\left(R_i^{(b)}(Z_i), T_i^{(b)}(Z_i) \right)$. The hazards specified in the main text lead to a simple imputation strategy based on Weibull random deviates. The resulting draws of $\left(R_i^{(b)}(Z_i), T_i^{(b)}(Z_i) \right)$ are compatible with $\bthetab$, $\gamma_i^{(b)}$, and $O_i$. 
If individual $i$ was censored before the nonterminal event occurred, we impute the missing event times according to the following algorithm.
\begin{enumerate}
\item Impute a candidate nonterminal event time $R^*$ from a Weibull distribution with shape parameter $\alpha^{Z_i, (b)}_1$ and scale parameter $\expb{-(\log(\gib \kappa^{Z_i, (b)}_1) + \bx_i'\bbeta_1^{Z_i,(b)})/\alpha^{Z_i}_1}$ that is truncated to have no mass below $C_i$.
\item Impute a candidate death time $T^*$ from a Weibull distribution with shape parameter $\alpha^{Z_i,(b)}_2$ and scale parameter $\expb{-(\log(\gib \kappa^{Z_i,(b)}_2) + \bx_i'\bbeta_2^{Z_i,(b)})/\alpha^{Z_i,(b)}_2}$ that is truncated to have no mass below $C_i$. If $T^* < R^*$, set $R_i(Z_i)^{(b)} = \Rbar$ and $T_i(Z_i)^{(b)} = T^*$. This gives us a complete $(R_i(Z_i)^{(b)}, T_i(Z_i)^{(b)})$ and the imputation process concludes. Otherwise, set $R_i(Z_i)^{(b)} = R^*$ and continue to Step~\ref{step:impsoj}.
\item Impute a sojourn time $S^*$ from a Weibull distribution with shape parameter $\alpha^{Z_i, (b)}_3$ and scale parameter $\expb{-(\log(\gib \kappa^{Z_i, (b)}_3) + \bx_i'\bbeta_3^{Z_i,(b)})/\alpha^{Z_i, (b)}_3}$. \label{step:impsoj}
\item Set the imputed death time $T_i(Z_i)^{(b)}$ to $R_i(Z_i)^{(b)} + S^*$.
\end{enumerate}
For individuals censored after the nonterminal event, the procedure starts at Step~\ref{step:impsoj} with the modification that the distribution of the sojourn time must be truncated to have no mass below $C_i - R_i(Z_i)$. After imputation, each individual has a complete set of four potential outcomes for all $B$ MCMC iterations.

\subsubsection{Imputation of counterfactual potential outcomes}\label{sec:impcounterfact}
Missingness in the outcome pair $\left(R_i(1-Z_i), T_i(1-Z_i) \right)$ is due to the so-called fundamental problem of causal inference. From assuming independent gamma-distributed frailties as in the main text, we have that posterior draws of $\left(R_i^{(b)}(1-Z_i), T_i^{(b)}(1-Z_i) \right)$ depend only on $\gamma_i^{(b)}$, $\btheta^{(b)}$, and $O_i$. 

Imputation of outcomes in the treatment arm counter to fact is actually simpler because there is no need to truncate so that the imputed values agree with the observed $(Y^R_i, \delta^R_i, Y^T_i, \delta^T_i)$. Replace $Z_i$ with $1 - Z_i$ and $C_i$ with 0, then follow the algorithm in Section~\ref{sec:impcens}.

\section{Discrepancy metrics}
Here we outline the algorithms for the calculation of posterior predictive $p$-values from two relevant discrepancy measures.

\subsection{Proportion always-alive}
\label{sec:discaa}
For $b = 1, \dots, B$ post-warmup draws of $\btheta$ from the posterior distribution and $i = 1,\dots, n$:
\begin{enumerate}
\item Draw frailty $\gamma_i^{rep (b)} \overset{i.i.d}{\sim} \mathrm{Gamma}\left(1/\sigma^{(b)}, 1/\sigma^{(b)}\right)$.
\item Set $Z_i^{rep} = Z_i^{obs}$, $C_i^{rep} = C_i^{obs}$, and $\bX_i^{rep} = \bX_i^{obs}$.
\item Simulate replicate uncensored factual potential outcomes $R_i^{rep (b)}(Z_i)$ and $T_i^{rep (b)}(Z_i)$ and counter-to-fact potential outcomes $R_i^{rep (b)}(1 - Z_i^{rep})$ and $T_i^{rep (b)}(1 - Z_i^{rep})$ using $\bX_i^{rep}, \bthetab$, and $\gamma_i^{rep (b)}$.
\item Apply censoring time $C_i^{rep}$ to obtain 
\[ \left(Y_i^{R, rep (b)}(Z_i^{obs}), \delta_i^{R, rep (b)}(Z_i^{obs}), Y_i^{T, rep (b)}(Z_i^{obs}), Y_i^{R, rep (b)}(1 - Z_i^{obs}), Y_i^{T, rep (b)}(1 - Z_i^{obs}) \right)\]
\item Determine principal states at $t_k$ for $k = 1, \dots, K$:
\[V_i^{rep(b)}(t_k) =
\begin{cases}
AA &
 Y_i^{T, rep (b)}(1) > t_k,
 Y_i^{T, rep (b)}(0) > t_k \\
 TK &
 Y_i^{T, rep (b)}(1) \leq t_k,
 Y_i^{T, rep (b)}(0) > t_k \\
 CK &
 Y_i^{T, rep (b)}(1) > t_k,
 Y_i^{T, rep (b)}(0) \leq t_k \\
 DD &
 Y_i^{T, rep (b)}(1) \leq t_k,
 Y_i^{T, rep (b)}(0) \leq t_k
\end{cases}
\] \label{step:pstate}
\item Calculate replicate fraction always-alive at time $t_k$, $\bar{V}_{AA, t_k}^{rep, (b)}$
\[ \bar{V}_{AA, t_k}^{rep, (b)}
=
n^{-1}
\sum_{i=1}^{n} 
\mathbb{1}\left( 
V_i^{rep (b)}(t_k) = AA
\right)
\]
\item Calculate corresponding fraction in the observed data set, $\bar{V}_{AA, t_k}^{obs, (b)}$
\[ \bar{V}_{AA, t_k}^{obs, (b)}
=
n^{-1}
\sum_{i=1}^{n} 
\mathbb{1}\left( 
V_i^{obs (b)}(t_k) = AA
\right)
\]
\item Calculate discrepancy measure $T_{AA, t_k}^{(b)} = \mathbb{1}\left(\bar{V}_{AA, t_k}^{obs, (b)} > \bar{V}_{AA, t_k}^{rep, (b)}\right)$.
\end{enumerate}
The posterior predictive $p$-value at $t_k$ for the discrepancy measure is $PPPV_{AA, t_k} = B^{-1} \sum_{b=1}^B T_{AA, t_k}^{(b)}$.

\subsection{Marginal survival within treatment arms}
The algorithm is identical to the one in Section~\ref{sec:discaa} through Step~\ref{step:pstate}. Then, for each $z \in \{0,1\}$ at a series of time points $t \in \{ t_1, \dots, t_K \}$:

\begin{enumerate}\setcounter{enumi}{4}
\item Using the observed data, calculate the marginal Kaplan-Meier survival estimate in arm $z$ at a grid of $K$ times $t_1,\dots,t_K$, where $t_K$ is the maximum observed event or censoring time in the observed data. For the $D^{obs}$ unique observed death times, let $\tau_m^{obs}$ be the $m^{th}$ observed ordered event time ($m = 1, \dots, D^{obs}$). Denote the number of deaths occurring at time $t$ in the $Z=z$ group by $d_z^{obs}(t)$, and the number at-risk at time $t$ in the $Z=z$ group by $r_z^{obs}(t)$.
\[ KM_{z}^{obs}(t_k) = \prod_{m : \tau_m^{obs} < t_k} \left[ 1 - \frac{d_z^{obs}(\tau_m^{obs})}{r_z^{obs}(\tau_m^{obs})} \right] \]
This does not depend on $\btheta$ and is therefore the same across all $b = 1,\dots, B$.
\item For the replicate data set corresponding to $b^{th}$ MCMC iteration, let $D^{rep, b}$ denote the unique replicate death times, let $\tau_m$ be the $m^{th}$ replicate event time ($m = 1, \dots, D^{rep, b}$).
\[ KM_{z}^{rep}(t_k) = \prod_{m : \tau_m < t_k} \left[ 1 - \frac{d_z^{rep}(\tau_m)}{r_z^{rep}(\tau_m)} \right] \]
\item Calculate discrepancy measure in $T_{KM,z,t_k}^{(b)} = \mathbb{1}\left( 
 KM_{z}^{obs}(t_k) >  KM_{z}^{rep}(t_k)
\right)$.
\end{enumerate}
The posterior predictive $p$-value at $t_k$ for the marginal survival discrepancy measure in group $z$ is $PPPV_{KM, z, t_k} = B^{-1} \sum_{b=1}^B T_{KM,z,t_k}^{(b)}$.

\subsection{Kolmogorov-Smirnov deviation from gamma frailty}
\begin{enumerate}\setcounter{enumi}{1}
\item Calculate the Kolmogorov-Smirnov (KS) test statistic as the maximum deviation of the empirical CDF of the replicate frailties $\hat{F}_n^{rep(b)}(x)$ from $F^{(b)}(x; \sigma^{(b)})$, the CDF of the gamma distribution with mean 1 and variance $\sigma^{(b)}$:
\begin{align*}
    KS^{rep(b)} = & \max_{x \in (0,\infty)} \left\vert \hat{F}_n^{rep (b)}(x) - F^{(b)}(x; \sigma^{(b)}) \right\vert \\
    = & 
    \max_{x \in (0,\infty)} \left\vert 
    n^{-1} \sum_{i=1}^n \left(\gamma_i^{rep(b)} < x \right)
    -
    \frac{\gamma\left(\sigmainv^{(b)}, \sigmainv^{(b)} x \right)}{\Gamma\left(\sigmainv^{(b)}\right)}
    \right\vert
\end{align*} where $\gamma\left(x; a\right)$ is the lower incomplete gamma function and $\Gamma(a)$ is the gamma function.
\item Calculate the corresponding KS statistic for the imputed frailties in the observed data:
\begin{align*}
    KS^{obs(b)}
    = & \max_{x \in (0,\infty)} \left\vert \hat{F}_n^{obs (b)}(x) - F^{(b)}(x; \sigma^{(b)}) \right\vert \\
    = & 
    \max_{x \in (0,\infty)} \left\vert 
    n^{-1} \sum_{i=1}^n \left(\gamma_i^{obs(b)} < x \right)
    -
    \frac{\gamma\left(\sigmainv^{(b)}, \sigmainv^{(b)} x \right)}{\Gamma\left(\sigmainv^{(b)}\right)}
    \right\vert
\end{align*}
\item Calculate discrepancy measure $T_{KS}^{(b)} = \mathbb{1}\left(
KS^{obs(b)} > KS^{rep (b)}\right)$.
\end{enumerate}
The posterior predictive $p$-value for the KS discrepancy measure is $PPPV_{KS} = 1 - B^{-1} \sum_{b=1}^B T_{KS}^{(b)}$. Subtracting from 1 ensures that $B^{-1} \sum_{b=1}^B T_{KS}^{(b)}$ closer to 1 (i.e., because of poor fit) correspond to $p$-values closer to 0, as with conventional frequentist $p$-values.

\section{Data application: frailties}

\begin{figure}[ht!]
\centering
\includegraphics[width=0.98\textwidth]{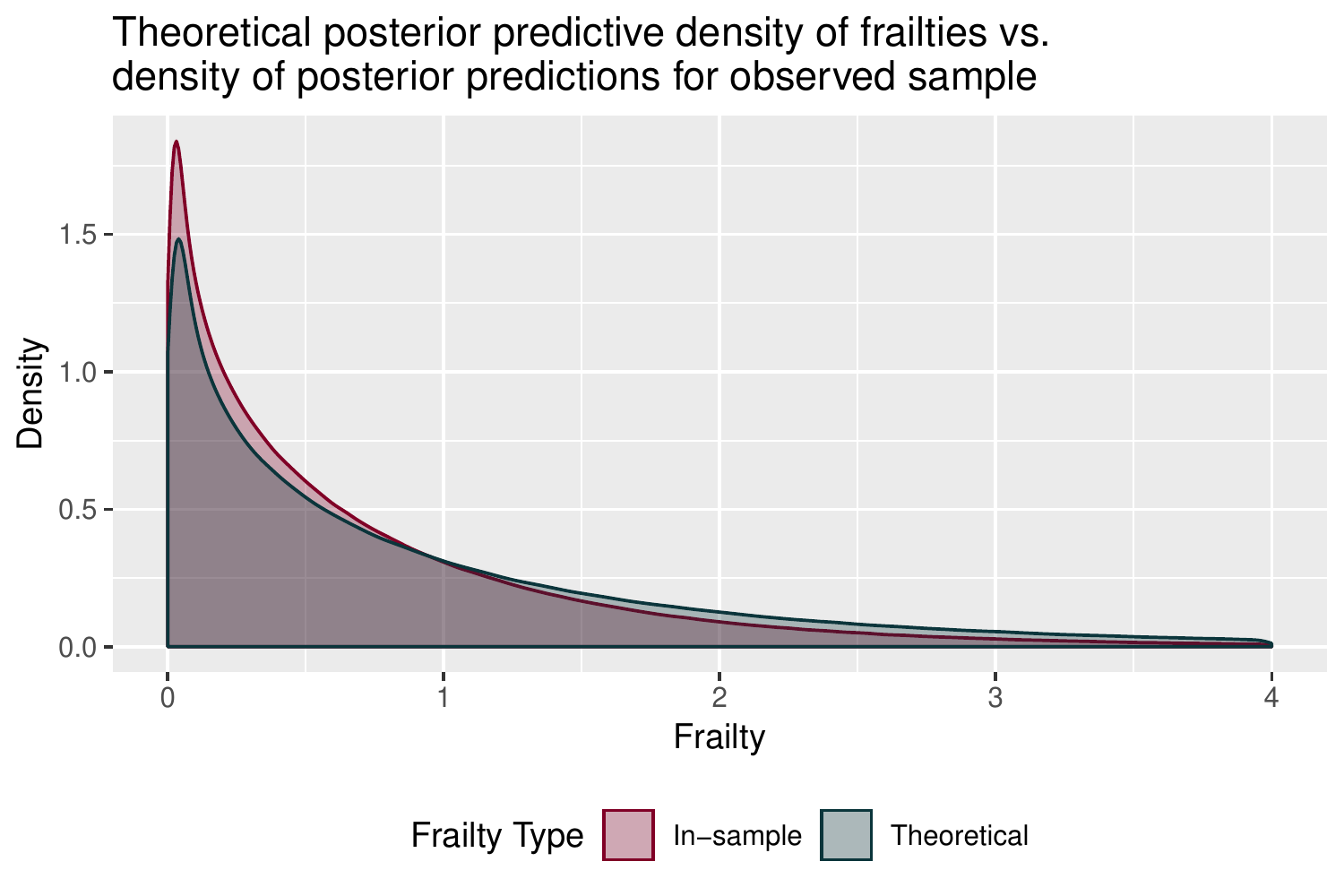}
\caption{Density of in-sample posterior predictive frailties compared to the samples from the posterior predictive distribution of latent frailties implied by $\sigma$} \label{fig:dafrailtydens}
\end{figure}

Figure~\ref{fig:dafrailtydens} shows density of posterior draws for $\bgamma$ for the \textcolor{dacolor}{6,840} in-sample individuals in red. The blue density shows the posterior predictive distribution implied by the posterior of $\sigma$ (i.e., the distribution of $\gamma_i$ for a new individual). The relative excess of frailties $< 1$ suggests that the individuals in the sample are healthier than expected by the model. This suggests that the gamma distribution may not be a good choice for the frailties, although this may be alleviated by more flexible specification of baseline hazards or covariate effects. Since the frailty is a driver of correlation in survival status across treatment arms, this form of misspecification for the frailty may explain the high $p$-values we observed for the $T_{AA}$ discrepancy metric.

\label{lastpage}

\end{document}